\documentclass[aps,showpacs,amsmath,amssymb,twocolumn,prl,floatfix,superscriptaddress]{revtex4}
\ifx\pdftexversion\undefined
\usepackage[dvips]{graphicx}
\else
 \usepackage[pdftex]{graphicx}
\fi
\usepackage{amsmath}
\begin{document}

% Use the \preprint command to place your local institutional report
% number in the upper righthand corner of the title page in preprint mode.
% Multiple \preprint commands are allowed.
% Use the 'preprintnumbers' class option to override journal defaults
% to display numbers if necessary
%\preprint{}

%Title of paper
\title{Dielectronic Resonance Method for Measuring Isotope Shifts }
%\title{Measuring the Hyperfine Shift in Cu-like $^{207}$Pb by Electron
%Scattering Resonances }
% repeat the \author .. \affiliation  etc. as needed
% \email, \thanks, \homepage, \altaffiliation all apply to the current
% author. Explanatory text should go in the []'s, actual e-mail
% address or url should go in the {}'s for \email and \homepage.
% Please use the appropriate macro for each each type of information
% \affiliation command applies to all authors since the last
% \affiliation command. The \affiliation command should follow the
% other information
% \affiliation can be followed by \email, \homepage, \thanks as well.
%\email[]{Your e-mail address}
%\homepage[]{Your web page}
%\thanks{}
%\altaffiliation{}
\author{R. Schuch}
\author{E. Lindroth}
\author{S. Madzunkov}
\author{M. Fogle}
\author{T. Mohamed}
\affiliation{Atomic Physics, Stockholm University, AlbaNova, S-106
91 Stockholm, Sweden}

\author{P. Indelicato}
\affiliation{
Laboratoire Kastler Brossel, \'Ecole Normale Sup\'erieure et
Universit\'e P. et M. Curie, Case 74, 4 place Jussieu, F-75252, Cedex 05,
France}

\date{\today}

\begin{abstract}
Longstanding problems in the comparison of very accurate
hyperfine-shift measurements to theory were partly overcome by
precise measurements on few-electron highly-charged ions. Still
the agreement between theory and experiment is unsatisfactory. In
this paper, we present a radically new way of precisely measuring
hyperfine shifts, and demonstrate its effectiveness in the case of
the hyperfine shift of $4s_{1/2}$ and $4p_{1/2}$ in
$^{207}\mathrm{Pb}^{53+}$. It is based on the precise detection of
dielectronic resonances that occur in electron-ion recombination
at very low energy. This allows us to determine the hyperfine
constant to around 0.6~meV accuracy which is on the order of 10\%.

\end{abstract}

% insert suggested PACS numbers in braces on next line
%31.30.Gs    Hyperfine interactions and isotope effects, Jahn-Teller effect
% OR 32.10.Fn     Fine and hyperfine structure ?
%34.80.Lx    Electron-ion recombination and electron attachment
%31.30.Jv    Relativistic and quantum electrodynamic effects in atoms
%and molecules?

\maketitle

The hyperfine structure (hfs) in atoms reflects fundamental
high-order effects in the interaction between the bound electrons
and the nucleus.  The energy shift contains contributions from
Quantum Electro-Dynamics (QED) and from the nuclear charge and
magnetic moment distribution \cite{gustavsson}. These effects have
been studied over decades by spectroscopy, mainly on neutral atoms
or singly charged ions. In such many-electron systems, the atomic
structure is complex. The theoretical predictions are further
hampered by imprecise knowledge of nuclear magnetic moments which,
if measured in a magnetic field, is {\em shielded} due to
diamagnetic effects~\cite{gustavsson}, which  can only be
estimated and corrected for through calculations.  Only recently,
experimental approaches to measure the hyperfine structure of
heavy highly-charged systems were opened \cite{klaft, lopez,
seelig, beiersdorfer}, giving  a new way to obtain nuclear moments
and to test QED in a new regime. Measurements were done for
several hydrogen-like ions: $^{209}\mathrm{Bi}^{82+}$,
$^{165}\mathrm{Ho}^{66+}$, $^{183,187}\mathrm{Re}^{74+}$, $^{203,
205}\mathrm{Tl}^{80+}$, and $^{207}\mathrm{Pb}^{81+}$\cite{klaft,
lopez, seelig, beiersdorfer}. In all cases the theory and
experiment do not agree in a satisfactory way. The largest
theoretical uncertainty comes from the unknown distribution of
nuclear magnetic moment (Bohr-Weisskopf effect), which is
complicated to predict due to nuclear
polarization~\cite{tomaselli}, but also precise knowledge of the
charge radius is crucial. On the other hand, a determination of
nuclear magnetic-moment distribution is impossible due to
uncertainties in the QED prediction. Although the measurements
with H-like ions are quite accurate ($2 \times 10^{-4}$ for
$^{207}\mathrm{Pb}$ \cite{seelig}), it is still essential to find
alternative approaches to measure hyperfine structure for
different ion charges where the above mentioned effects contribute
differently.

Here we demonstrate a first measurement of the hyperfine structure
in $^{207}\mathrm{Pb}$ by observing dielectronic recombination
(DR) resonances for both $^{208}\mathrm{Pb}^{53+}$ (zero nuclear
spin) and $^{207}\mathrm{Pb}^{53+}$ (nuclear spin one-half). Using
the two Pb isotopes in charge state 53+, one clearly observes
hyperfine effects in $4s_{1/2}$ and $4p_{1/2}$ F=1 and F=0 states.
The importance of this observation is that it provides means to
easily measure the hyperfine interaction of
$^{207}\mathrm{Pb}^{53+}$, and other ions with a different number
of electrons, as compared to what is currently possible. By
measuring the same nucleus with different numbers of electrons one
can measure their influence on the magnetic interaction and check
the theories including QED and probably do even better by
extrapolating to low electron numbers. A similar idea, combining
the $1s$ hyperfine shift in H-like and the $2s$ shift in Li-like
ions to extract QED contributions, was recently proposed in a
theoretical paper\cite{shabaev}. A further application of this
method can be found in the new accumulator rings for radioactive
nuclei to measure their nuclear magnetic properties by low-energy
electron scattering in the cooler\cite{henn}.

The spectroscopic principle of using DR resonances is that the
valence electron is excited by the capture of a free electron into
a Rydberg state. The measured resonances are near threshold, thus
the Rydberg energy must balance to a large part the excitation
energy of the valence electron. Higher order effects in the
interaction of the Rydberg electron with the core electrons are
weak and can be estimated to excellent accuracy. The measurement
is checked by comparing the DR resonances with those of
$^{208}\mathrm{Pb}^{53+}$ where no hyperfine splitting exists. In
the $^{208}\mathrm{Pb}^{53+}$ case we reached before an accuracy
that enabled us to obtain spectroscopic data for stringent tests
of many-electron QED effects in advanced atomic structure
calculations\cite{lindroth}.

Selected $^{207}\mathrm{Pb}$ or $^{208}\mathrm{Pb}$ isotopes are
injected as singly charged ions from an isotope separator, at the
Manne Siegbahn Laboratory in Stockholm, into an electron-beam ion
source, where the highly charged Pb ions are created. From there
they are injected into the heavy-ion storage ring CRYRING, and
accelerated to several MeV/amu. After electron cooling the ion
beam for 1 s, the recombination rates as a function of relative
energy between electrons and ions are measured. This is achieved
by ramping the electron energy up and down, from the cooling
condition, to a maximum energy in the center-of-mass frame so that
recombination spectra are obtained both with the electrons being
slower and faster than the ions. Thus, we can check the space
charge of the electron beam and the longitudinal drag force
exerted on the ions by the electrons during detuning and make
appropriate corrections \cite{zong, madzunkov}. These corrections
are important in order to obtain accurate energy scales in the
spectra. The energy calibration was checked thoroughly and found
to be accurate to within 0.1 meV at energies $\leq 0.3$ eV. For
these low energies one takes full advantage of the small energy
spread of the adiabatically expanded electron beam \cite{danared}
and obtains a resolution on the order of 1 meV. From fits to the
resonances, we confirm the longitudinal and transversal
temperatures of the electron beam to be k$T_l=0.1$~meV and
k$T_t=2$~meV, respectively.

\begin{figure}
\includegraphics[width=\columnwidth]{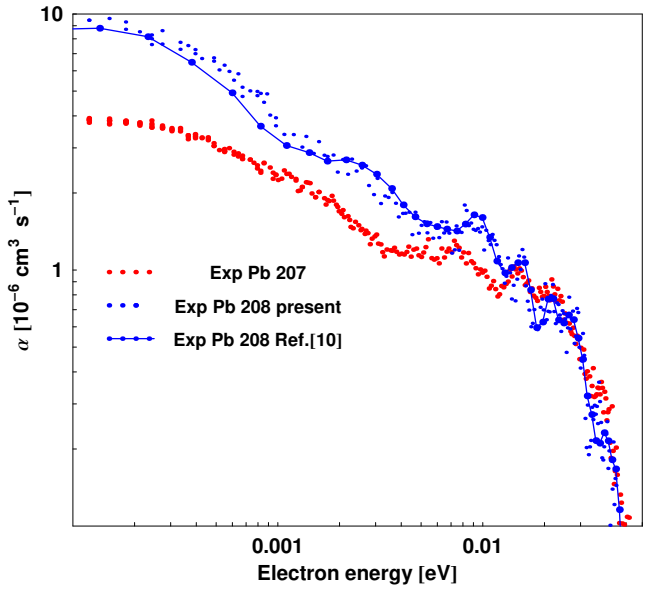}
 \caption{Measured  dielectronic
recombination spectra with $^{207}\mathrm{Pb}^{53+}$ and with
$^{208}\mathrm{Pb}^{53+}$.  \label{fig1}}
\end{figure}

The consistency of the measurement is checked by switching between
the $^{207}\mathrm{Pb}^{53+}$ and $^{208}\mathrm{Pb}^{53+}$ beams.
Small changes in the magnet field strengths were made for the two
isotope ions to have the same velocity (determined by electron
cooling) and the same circulation frequency in the ring (measured
by the Schottky frequency). We also compare the
$^{208}\mathrm{Pb}^{53+}$ data measured in this run with those
measured earlier (with k$T_l=0.08~$meV and k$T_t=1$~meV) and
published in Ref.~\cite{lindroth} (see Fig.~\ref{fig1}).
Considering the small temperature difference, these data sets
agree within the statistical error(its size is seen in the spread
of the points). In Fig.~\ref{fig1} we also show the recombination
rate coefficient that was obtained with $^{207}\mathrm{Pb}^{53+}$
stored in the ring. The difference between the two sets of
experimental spectra must then be due to the isotope effect, in
particular the hyperfine shifts of the $F=0$ and 1 states in
$4s_{1/2}$ and $4p_{1/2}$ of $^{207}\mathrm{Pb}^{53+}$.

We have evaluated the isotope dependent contributions to
Pb$^{53+}\left(4\ell_{1/2} \right)$ using both Relativistic
Many-Body Perturbation Theory (RMBPT)~\cite{amp} and a
Multi-Configuration Dirac-Fock (MCDF) computer
package~\cite{desclaux,boucard}. We have carefully checked the
agreement between the methods. The dominating effect, listed in
detail on the first six lines of Table~\ref{tab:hfs}, is due to
the coupling between the total electronic angular momentum and the
nuclear magnetic moment of the spin one half $^{207}$Pb nucleus.
Most of the contribution is accounted for already with a
single-configuration DF calculation using a point nucleus, as seen
on the first line. The lead wave function has then been evaluated
in a potential from an extended nucleus, which results in a $\sim
10$~\% change in the $4s$ hyperfine structure, as seen on the
second line of Table~\ref{tab:hfs}. We use a Fermi distribution of
the nuclear charge with thickness parameter 2.3 fm and a half
density radius adjusted to produce a charge root mean square
radius (rms) of $5.4942\,(0.0013)$ fm~\cite{angeli}. The wave
function has also been allowed to adjust to the Breit interaction
between the electrons and to the Uehling potential (responsible
for the bulk of the vacuum polarization), but the effect on the
hyperfine structure is very small. All these contributions agree
within the displayed digits between the two computational methods.
The polarization of the closed shell core by the valence electron
give rise to important corrections to the hyperfine structure.
These effects can only be accounted for when spherical symmetry is
abandoned, but they are still not true correlation effects. We
have calculated this core-polarization using RMBPT(fourth line).
The dominating part of the core-polarization is accounted for
already in second order (one order in the hyperfine structure and
one order in the Coulomb interaction). The result change with only
a few $\mu$eV when higher order core polarization is taken into
account. Since also true correlation, calculated with either RMBPT
or MCDF,  gives negligible contributions, see Table~\ref{tab:hfs},
it is evident that the many-body aspect of Pb$^{53+}$ does not
limit the potential accuracy of the determination of the hyperfine
structure. An approximate Bohr-Weisskopf effect is included,
assuming the same mean spherical radius for the magnetic moment
distribution as was used for the charge distribution, following
e.g. Ref.~\cite{boucard}.  Also here the two computational methods
agree to within one unit in the last quoted digit.

An additional isotope dependent contribution originate from the
slightly different potentials produced by nuclei with different rms
(volume shift). According to Angeli~\cite{angeli} the rms of
$^{207}$Pb is $0.0068\,(0.0002)$ fm smaller than that of $^{208}$Pb.
This leads to $\sim 1.6$ meV larger binding energy for the $4s$
state of $^{207}$Pb, corresponding to
 $\sim 0.2$ meV $/10^{-3}$ fm, as seen in Table~\ref{tab:hfs}.

The experimental evaluations of the magnetic moment of e.g.,
$^{207}$Pb have been critically evaluated in
Ref.\cite{gustavsson}, indicating that the true value should be
within error bars of the Atomic Beam Magnetic Resonance (ABMR)
result by Brenner  $\mu_I = 0.5918\, (14)\mu_N$~\cite{brenner} and
the Nuclear Magnetic Resonance (NMR) result by Lutz and
Stricker~\cite{lutz}, quoted in Ref.\cite{gustavsson} as $\mu_I =
0.5925\, (6)\mu_N$. While the optical pumping result by Gibbs and
White~\cite{gibbs}, which is $\approx 1.3\%$ below
Ref.~\cite{brenner}, seems to suffer from systematic errors. Here
we have used the value from Ref.~\cite{brenner}.

As in $^{208}\mathrm{Pb}$~\cite{lindroth} the recombination goes
through resonances in  $\mathrm{Pb}^{52+}$,
$\mathrm{Pb}^{53+}\left(3d^{10}4s_{1/2}\right) + e^- \rightarrow
\mathrm{Pb}^{52+}\left(3d^{10}4p_{1/2}18 \ell_j \right)_J$. The term
splitting of these excited states range from a few meV for the
states around the ionization threshold, $j=21/2$, of
$^{208}\mathrm{Pb}^{52+}$\cite{lindroth} to practically zero for the
highest angular momenta. When considering hyperfine structure we
note first that the J- dependent term splitting and the hyperfine
splitting of the $4p_{1/2}$ orbital energy are of the same order of
magnitude. The energy levels are then expected to be governed by
intermediate coupling and will
 not scale linearly with the hyperfine shift; each level will be an
 individual mix of the two hyperfine components of the inner
 electron orbital. The hyperfine
interaction involving the outer $18 \ell_j$ electron will be much
smaller due to the $1/n^3$ scaling of the radial matrix elements
and will be neglected in the following. Starting from the
$\left(4p_{1/2}18 \ell_j \right)_J$ energy levels in the absence
of hyperfine interaction~\cite{lindroth}, using the assumption
that only terms (J) belonging to the same configuration will mix
due to the hyperfine interaction, and using the knowledge of the
splitting of the $4p_{1/2}$ in the absence of the Rydberg
electron, it is straight forward to generate the hyperfine
perturbed resonance energies with standard angular momentum
theory. Without hyperfine interaction each $4p_{1/2}18 \ell_j$
configuration forms two $J$-terms. With hyperfine interaction we
get four energy levels of which two mix the $J$-terms. The
recombination cross section can now be constructed from the energy
positions relative the initial state of $\mathrm{Pb}^{53+}$, the
autoionization rates and the stabilizing radiative rates to bound
states of $\mathrm{Pb}^{52+}$. The autoionization rate varies
strongly and is governed by the actual mixture just discussed. The
radiative rate is for these high angular momenta states dominated
by the probability for the $4p \rightarrow 4s$ transition and is
thus slowly varying and nearly constant for all the investigated
resonances. The position of a resonances is affected by the
positions of the hyperfine perturbed excited states as  well as of
the shifted initial state. The ground state of $\mathrm{Pb}^{53+}
4s$ is with hyperfine interaction split into one state with F=0
and one with F=1. The F=1 state has higher energy, but its decay
to F=0 is very slow. The expected life time is $ 2.3\times10^{4}$
s \cite{shabaev} which is much longer than the time from ions
production to the measurement($\sim$ 5s). We have thus assumed a
 statistical population of the two $\mathrm{Pb}^{53+} \left( 4s_{1/2}\right)_F$ states
 and the recombination cross section from the two possible initial
 states have been calculated and added with statistical weights.
This cross section has subsequently been folded with the electron
 beam velocity distribution mentioned above.

\begin{figure}
\includegraphics[width=\columnwidth]{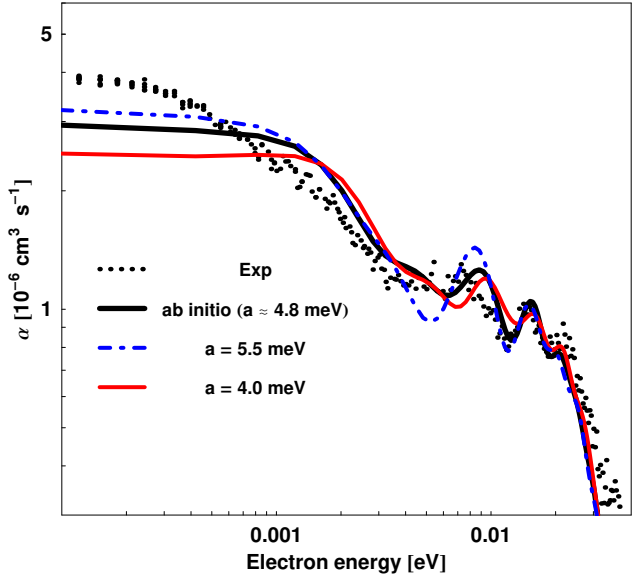}
\caption{Calculated recombination cross sections for
$^{207}\mathrm{Pb}^{53+}$ with different values of the hyperfine
constant, $a$, as explained in the text.\label{fig2}}
\end{figure}

\begin{figure}
\includegraphics[width=\columnwidth]{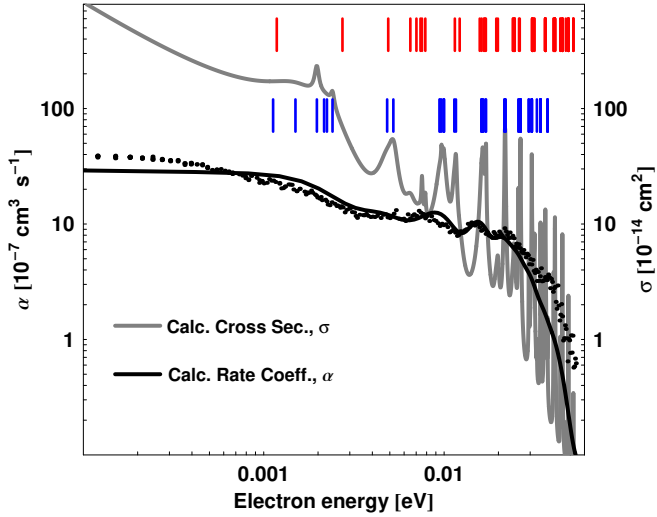}
\caption{Calculated resonance positions for initial states
$^{207}\mathrm{Pb}^{53+}\left ( 4s_{1/2} \right)_{F=0} $, upper
 sequence of levels, and $^{207}\mathrm{Pb}^{53+}\left (
4s_{1/2}\right)_{F=1}$, lower sequence of levels. The grey line
shows the cross section assuming statistical population of these
two hyperfine structure components. From this cross section the
rate coefficient, the black solid line, is generated. It should be
compared to the measured rate coefficient for
$^{207}\mathrm{Pb}^{53+}$, shown as black dots. \label{fig3}}
\end{figure}

 The nuclear spin dependent part of the hyperfine energy shift is
$\Delta E_{HFS}=a/2[F(F+1)-I(I+1)-J(J+1)]$, where $a$ is the
hyperfine splitting constant for a given state, $J$ is the
electronic  angular momentum, $I$ is the nuclear spin, and $F$ is
the total angular momentum. Using the fact that $3a_{4p} \approx
a_{4s}$ to within $2 - 3$\%, as shown in Table~\ref{tab:hfs}, we
generate a synthetic spectrum derived from the resonance
parameters as described above, and test the sensitivity with
respect to small variations in $a=a_{4p}$. The theoretical shifts
in Table~\ref{tab:hfs} are approximately reproduced by $a=4.8$
meV. Fig~\ref{fig2} shows the spectrum generated with the
theoretical shifts as well as with $a$ values deviating from it
with around $\pm 0.7$ meV. The spectrum generated from the
calculated shifts seems to fit the data best. Note that different
peaks in the spectrum are rather differently affected when $a$ is
changed, reflecting the non-linear shift of the resonances. With
$a = 5.5$ meV the synthetic spectrum shows much more pronounced
resonances and with $a = 4.0$ meV the dip around $0.01$ eV has
nearly disappeared. We estimate from this procedure that with the
present energy spread of the electron beam we can get an accuracy
in determining $a$, from a fit to the resonances, to the order of
$\pm 0.3$ meV. Here the statistical error was taken into account.
With a specially designed electron target of lower energy spread,
one could reduce the uncertainty further. Fig~\ref{fig3} shows the
positions of all the resonances and the cross section before
folding with the electron temperature.

At the present stage we are also limited in our accuracy by the
uncertainty in describing the resonances. This problem already
exists in the $^{208}\mathrm{Pb}^{53+}$ case~\cite{lindroth}.
There is a so far unexplained discrepancy between theory and
measured rate coefficient at energies below 1 meV. That may cause
the systematic uncertainty to reach 1 meV. Several resonances
above 1 meV are well described by the calculated series. Another,
unexplained discrepancy appears at energies larger then 0.03 eV
(see Fig. 3). As can be seen from Fig. 3 the highest energy
resonances contribute insignificantly to the cross section. The
reason might be that the autoionization rates approach zero for
these high angular momenta states, but even small electromagnetic
fields in the cooler could change this by inducing a mixture with
states of low angular momentum. This possibility has to be
investigated further, but will not affect the conclusions here
since in this range one is not very sensitive to a hyperfine
splitting in the order of meV.

In conclusion, the hyperfine splittings in $4p_{1/2}$ and
$4s_{1/2}$ state of the Cu-like $^{207}\mathrm{Pb}$ isotope were
measured by a new method to an accuracy of the order 10\% of the
hfs constant. Instead of the generally applied optical
spectroscopy, the splitting was observed in dielectronic
resonances in low-energy electron-ion recombination. This opens
new possibilities to measure this important physical quantity in a
new energy range and ionization stage for determining the
different contribution to its value.

We thank the CRYRING crew for the skillful operation of the storage
ring. We acknowledge support from the Swedish research council (VR).
Laboratoire Kastler Brossel is Unit{\'e} Mixte de Recherche du CNRS
n$^{\circ}$ 8552.

%*********FIGURE 1***********

%*********FIGURE 2***********

%*********FIGURE 3***********

\begin{table}[htb]
\squeezetable
 \caption{ \label{tab:hfs} Binding energy of $^{207}$Pb$^{53+} \left(4j_{1/2} \right)_F$ relative the
 $\left( 4j_{1/2} \right)$ states
 in $^{208}$Pb$^{53+}$ (meV).
}
\begin{ruledtabular}
\begin{tabular}{lrrrr}
 & \multicolumn{2}{c}{$4s_{1/2}$} & \multicolumn{2}{c}{$4p_{1/2}$}  \\
& F=0 &  F=1  & F=0  & F=1 \\
 \hline
Contributions $\sim \mu_I$\footnotemark[1] \\
\hline
point size $ \mu_I$\\
 Dirac-Fock point nucleus &    -12.16  &   4.05    &   -3.66   &   1.22        \\
 $\Delta$ nuclear size &    1.37    &   -0.46   &   0.12    &   -0.04       \\
 $\Delta$ Breit and Vac. pol. & -0.03   &   0.01    &   0.03    &   -0.01       \\
 $\Delta$ core-polarization\footnotemark[2]  &    -0.34   &   0.11    &   -0.12   &   0.04        \\
  $\Delta$ correlation &    -0.03   &   0.01    &   0.04    &   -0.01       \\
 distributed $\mu_I$                                    \\
 (Bohr-Weisskopf effect) &  0.23    &   -0.08   &   0.02    &   -0.01       \\
                                    \\
Sum &   -10.97  &   3.66    &   -3.57   &   1.19        \\
\hline
Contributions independent of $\mu_I$                                    \\
\hline
Volume shift &  -1.59   &   -1.59   &   -0.08   &   -0.08       \\
                                    \\
Total  &    -12.56  &   2.06    &   -3.65   &   1.11        \\

\end{tabular}
\end{ruledtabular}
\footnotetext[1] {$\mu_I/\mu_N = 0.5918$, Ref.~\cite{brenner}} \\
\footnotetext[2] {Includes polarization due to both the Coulomb and
the Breit interaction, the latter contributes with 10-14\% }
\end{table}

\begin{thebibliography}{}
\bibitem{gustavsson}
M.~G.~H.~Gustavsson and A--M.~M{\aa}rtensson--Pendrill, Phys. Rev.
A \textbf{58}, 3611 (1998).
\bibitem{klaft}
I.~Klaft \emph{et al.}, Phys. Rev. Lett. \textbf{73}, 2425 (1994).
\bibitem{lopez}
J.~R.~Crespo~L\'{o}pez-Urrutia \emph{et al.}, Phys. Rev. Lett.
\textbf{77}, 826 (1996).
\bibitem{seelig}
P.~Seelig \emph{et al.}, Phys. Rev. Lett. \textbf{81}, 4824
(1998).
\bibitem{beiersdorfer}
P.~Beiersdorfer \emph{et al.}, Nucl. Instr. Meth. B \textbf{250},
62 (2003).
\bibitem{tomaselli}
M.~Tomaselli \emph{et al.}, Phys. Rev. C \textbf{58}, 1524 (1998).
\bibitem{shabaev}
V. M. Shabaev \emph{et al.}, Phys. Rev. A \textbf{57}, 149-156
(1998).
\bibitem{henn} Intern. Accel. Facility for Beams of Ions and Anti-Protons,
W.F. Henning (eds.), GSI Darmstadt, Germany, 2001, http://www.
gsi.de/GSI-future/cdr/ and R. Schuch and T. Stoehlker, Nucl.
Instr. Meth. to be publ.
\bibitem{gibbs}H.M. Gibbs and C.M. White, Phys. Rev. 188, 180 (1969)
\bibitem{lindroth}E.~Lindroth \emph{et al.}, Phys. Rev. Lett. \textbf{86}, 5027
(2001).
\bibitem{zong}W. Zong \emph{et al.}, Phys. Rev. A \textbf{56}, 386,  (1997)
\bibitem{madzunkov}S.~Madzunkov \emph{et al.}, Phys. Rev. A \textbf{65}, 032505
(2002).
\bibitem{danared}
H.~Danared \emph{et al.}, Nucl. Instr. Meth. A{\textbf 441}, 123
(2000).
\bibitem{amp} A--M.~M{\aa}rtensson--Pendrill and A. Ynnerman Phys. Scr. \textbf{41}, 329 (2000).
\bibitem{desclaux} J. P. Desclaux.\textsl{A Relativistic Multiconfiguration Dirac-Fock Package}.
in Meth. and Techn. in Comput. Chem. Clementi, E. Ed. STEF, 1993.
\bibitem{boucard}S. Boucard, P. Indelicato, Eur. Phys. J. D\textbf{8}, 59 (2000).
\bibitem{angeli}I. Angeli, At. Data Nucl. Data Tables \textbf{87}, 185 (2004).
\bibitem{brenner} T. Brenner, PhD. Thesis, Inst. f. Angew. Physik, Univ. Bonn, 1988
\bibitem{lutz} O. Lutz and G. Stricker, Phys. Lett, \textbf{35 A}, 397 (1971).
\bibitem{tokman}M.~Tokman \emph{et al.}, Hyp. Int. \textbf{132}, 285 (2001).
\bibitem{fire}R.B. Firestone, Table of Isotopes 8th ed. Wiley, New York (1996)
\end{thebibliography}
\end{document}